\documentclass[twoside]{article}

\usepackage{PRIMEarxiv}

\usepackage[utf8]{inputenc} 
\usepackage[T1]{fontenc}    
\usepackage{hyperref}       
\usepackage{url}            
\usepackage{booktabs}       
\usepackage{amsfonts}       
\usepackage{nicefrac}       
\usepackage{microtype}      
\usepackage{lipsum}
\usepackage{fancyhdr}       
\usepackage{graphicx}       
\graphicspath{{media/}}     
\usepackage{amsmath} 
\usepackage{float}

\usepackage{enumitem}
\usepackage{soul}
\usepackage{multirow}
\usepackage{multicol}
\usepackage{xcolor}
\usepackage{array}

\definecolor{darkgreen}{rgb}{0,0.7,0}

\title{Toward Robust URL Extraction for Open Science: \\A Study of arXiv File Formats and Temporal Trends}

\author{
  Rochana R. Obadage \\
  Old Dominion University  \\
  Norfolk, VA, USA \\
  \texttt{rochana@cs.odu.edu} \\
  \And
  Lamia Salsabil  \\
  Old Dominion University  \\
  Norfolk, VA, USA \\
  \texttt{lsals002@odu.edu} \\
  \And
  Sawood Alam  \\
  Internet Archive  \\
  San Francisco, CA, USA \\
  \texttt{sawood@archive.org} \\
  \And
  Bipasha Banarjee  \\
  Virginia Tech  \\
  Blacksburg, VA, USA \\
  \texttt{bipashabanerjee@vt.edu} \\
  \And
  William A. Ingram  \\
  Virginia Tech  \\
  Blacksburg, VA, USA \\
  \texttt{waingram@vt.edu} \\
  \And
  Edward A. Fox  \\
  Virginia Tech  \\
  Blacksburg, VA, USA \\
  \texttt{fox@vt.edu} \\
  \And
  Jian Wu \\
  Old Dominion University  \\
  Norfolk, VA, USA \\
  \texttt{j1wu@odu.edu} \\
}

\begin{document}
\maketitle

\begin{abstract}

In this work, we study how URL extraction results depend on input format. We compiled a pilot dataset by extracting URLs from 10 arXiv papers and used the same heuristic method to extract URLs from four formats derived from the PDF files or the source LaTeX files. We found that accurate and complete URL extraction from any single format or a combination of multiple formats is challenging, with the best F1-score of 0.71. Using the pilot dataset, we evaluate extraction performance across formats and show that structured formats like HTML and XML produce more accurate results than PDFs or Text. Combining multiple formats improves coverage, especially when targeting research-critical resources. We further apply URL extraction on two tasks, namely classifying URLs into open-access datasets and software and the others, and analyzing the trend of URLs usage in arXiv papers from 1992 to 2024.  These results suggest that using a combination of multiple formats achieves better performance on URL extraction than a single format, and the number of URLs in arXiv papers has been steadily increasing since 1992 to 2014 and has been drastically increasing from 2014 to 2024. The dataset and the Jupyter notebooks used for the preliminary analysis are publicly available at \textit{\url{https://github.com/lamps-lab/arxiv-urls}}.
\end{abstract}


\section{Introduction}

Recent research found that accessibility of datasets and software is crucial for reproducing the research findings of a paper \cite{kenny-icdar-2023}. As a digital library, arXiv \cite{arxiv, ginsparg2011yearsagotoday}, hosting over 2.3 million open-access full-text papers, serves as a critical platform for early research dissemination and long-term archiving. However, the digital resources these papers reference face a troubling reality: they disappear at an alarming rate \cite{url_decay, link_rot}. This link rot undermines the accessibility of shared datasets and software, which can further play a negative impact on reproducibility of research findings.  

\begin{figure}[ht]
\vspace{-0.1cm}
  \centering
    \setlength{\fboxsep}{4pt} 
    \setlength{\fboxrule}{0.5pt} 
  \fbox{\includegraphics[width=0.75
  \linewidth]{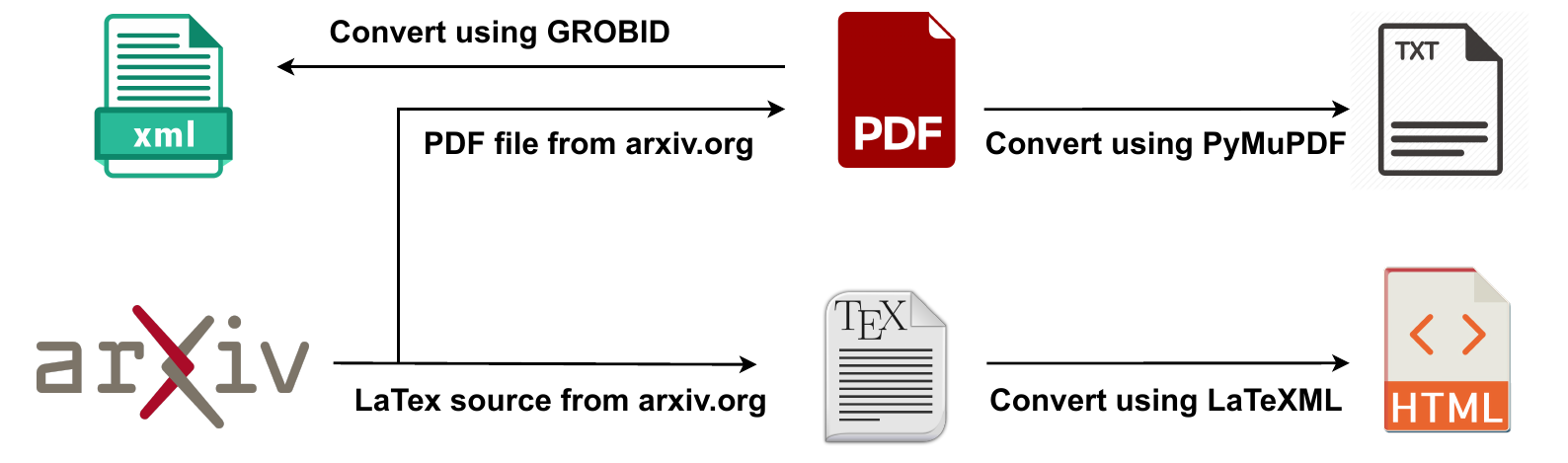}}
  \caption{Different file formats for an arXiv paper and how they are obtained for our study.}
  \label{fig:arxiv_file_formats}
\vspace{-0.3cm}
\end{figure}

Extracting URLs from PDF documents is an important task for a variety of downstream tasks such as URL classification, meta-analysis, and web archiving. Most existing methods first extract Text from PDFs \cite{pdf-text-1, pdf-text-2, pdf-text-3} and then use heuristic methods to identify URLs. However, this intuitive method  has several drawbacks. First, mainstream text extraction tools do not preserve the text flow in the original paper. This leads to truncated lines and thus URLs. Second, URLs may appear in the text and annotation layers of PDFs but most text extractors neglect URLs in annotation layers. Finally, in certain tasks, it is necessary to identify the locations of URLs, such as the footnotes or references, but it is very challenging to use a heuristic method (such as regular expressions) to accurately identify the locations.

arXiv provides papers in multiple formats including PDF and LaTeX source. Recently, arXiv launched an initiative to use HTML to publish research papers. The authors first submit the source files in LaTeX, which are then automatically converted to HTML format. However, the complexities of conversion may introduce noise or artifacts that prevent URLs from being accurately extracted. Compared with PDFs, HTML documents are more machine readable and thus much easier to extract URLs. In addition, GROBID is an open-source machine learning library designed to extract, parse, and re-structure scholarly papers. Its output is in TEI format, a standard for encoding and representing texts in scholarly papers. The output of GROBID marks rich semantic information such as metadata, references and segmented sections. How the performance of URL extraction depends on the input document format has not been well investigated. Our work aims to fill this gap using arXiv papers.

arXiv is a server that allows authors to submit and publish preprint papers. Since its launch in 1991, this digital service provided by arXiv has been extremely popular across a variety of scientific disciplines, and several sister preprint servers for fields not covered by the original arXiv, such as bioRxiv and medRxiv, have been established afterward. Understanding the dependence on the input format will help us to improve the performance of URL extraction. The comparison results will also guide us to develop data-ensemble methods that aggregate extraction results from different inputs. We systematically evaluate URL extraction performance across four full-text formats (Figure \ref{fig:arxiv_file_formats}) and analyze temporal trends in URL usage patterns from 1992 to 2024. 

Our contributions are twofold. First, we provide a comparative evaluation of URL extraction performance across four file formats, identifying which formats yield the best performance. Second, we present a longitudinal analysis of URLs in arXiv papers, revealing how the overall usage of URLs in a yearly sample has evolved over three decades. 

\section{Related Work}

Researchers have explored various methods for extracting URLs from scholarly documents, but most approaches focus on a single file format \cite{pdf-text-1, pdf-text-2, pdf-text-3}. Large-scale efforts like S2ORC \cite{lo-wang-2020-s2orc} and Semantic Scholar \cite{kinney2025semanticscholaropendata} mainly extract URLs from PDFs or preprocessed Text using regular expressions and heuristic filters to identify valid web addresses. Tools such as GROBID \cite{GROBID} and PyMuPDF \cite{PyMuPDF} are designed for PDF-based extraction, targeting structured information and references. GROBID uses PDFBox to extract Text from PDFs before applying ML models to parse the document. 

Some recent studies demonstrated the benefits of using HTML or XML formats, because these preserve more semantic structure \cite{Nasar, lo-wang-2020-s2orc} in the original document and are more ``machine-friendly'', to retain the fidelity of the original content. However, these studies do not focus specifically on URL extraction and do not perform direct comparisons across multiple formats as input. 

Studies show that links in academic papers decay over time, with rates varying by domain and hosting platform. Klein et al. documented widespread link rot \cite{10.1371/journal.pone.0115253}, and Hennessey et al. found that some domains are more vulnerable than others \cite{Hennessey2013}. A 2025 study of over 12,000 GitHub repositories found that more than 10\% had no archived version and that even archived pages often suffered from damage or missing source files \cite{10.1145/3717867.3717920}.
Escamilla et al. (2023) showed that a hybrid classifier can identify open-access data and software URIs across both major and niche platforms \cite{escamilla2023itsjustgithubidentifying}. However, these works either focus on particular domains, e.g., Git Hosted Platforms (GitHub, GitLab, etc.) or use off-the-shelf software without thorough evaluation. 

The current literature has two key gaps. First, most URL extraction tools perform information extraction based on text converted from PDFs and do not compare performance across different formats. As we will show, this may miss a significant number of URLs due to the broken content flow when text is extracted from PDFs. Second, there are no benchmarks designed specifically for extracting URLs from scholarly papers. “Hence, it is difficult to assess or improve how well a system captures URLs extracted from various formats. 

\section{Dataset} \label{sec:dataset}

To evaluate the URL extraction performance, we must build a dataset containing research papers in multiple formats (PDF, Text, XML, and HTML). We constructed our pilot dataset from arXiv (version: January 2024), which contains over 2.3 million open-access full-text research papers since 1991. We used stratified random sampling to select 1,161 papers, drawing three papers from each ``year, month'' stratum from 9107 to 2401. We applied \texttt{PyMuPDF} (v1.24.13) \cite{PyMuPDF} to extract Text directly from PDFs. Among the sampled papers, 726 had LaTeX source files. Only 17 included HTML versions from arXiv, hence we used the external tool \texttt{LaTeXML} (v0.8.8) \cite{githubGitHubBrucemillerLaTeXML} to convert LaTeX source files to HTML, which is the same tool used by arXiv to generate HTML files. Out of 726 LaTeX source files, the tool successfully converted 204 into HTML. The remaining files could not be converted due to inherent limitations and conversion errors in the tool.

\vspace{-2pt}

\renewcommand{\arraystretch}{1.3} 
\begin{table}[H]
\small
\centering
\caption{File format coverage and conversion tools used for the randomly selected sample of 1,161 arXiv papers}
\label{tab:format_coverage}
\begin{tabular}{p{0.13\columnwidth}p{0.14\columnwidth}p{0.12\columnwidth} p{0.25\columnwidth}}
\hline
\textbf{Format} & \textbf{Conversion Tool} & \textbf{\hfill \# Papers} & \textbf{\hfill \# Papers with extracted URLs} \\ \hline

PDF & --- & \hfill 1,161 & \hfill --- \\ \hline
Text & PyMuPDF \cite{PyMuPDF} & \hfill 1,161 & \hfill 260 \\ \hline
LaTeX & --- & \hfill 726 & \hfill 252 \\ \hline
HTML & LaTeXML \cite{githubGitHubBrucemillerLaTeXML} & \hfill 204 & \hfill 134 \\ \hline
XML & GROBID \cite{GROBID} & \hfill 60 & \hfill 60 \\ \hline
\end{tabular}
\end{table}

\vspace{-8pt}
After collecting and converting the files, we automatically excluded papers that do not contain any URL in any format. Table \ref{tab:format_coverage} summarizes the number of papers for each format and the number of papers from which we were able to extract at least one URL. Although all 1,161 papers were available in both PDF and Text form, 
only 60 papers contained URLs in all three formats: Text, LaTeX, and HTML. We selected this subset and used \texttt{GROBID} (v0.8.1) \cite{GROBID} to convert their PDFs into XML format. Finally, from the 60 papers that contained URLs in all five formats, we randomly selected 10 papers as the pilot dataset for our preliminary evaluation. The pipeline to generate different formats is shown in Figure \ref{fig:arxiv_file_formats}.

\section{URL Extraction Performance}

\subsection{Ground Truth}

To build the ground truth, we manually inspected the PDF versions of 10 selected papers (see Section \ref{sec:dataset}) and identified all valid URLs (ground truth). In our study, we consider a URL to be \textbf{valid} if it appears in the PDF version. This process resulted in a ground truth set of 87 valid URLs.

As a preliminary study, we apply a heuristic method for all file formats and try to identify URLs using markups of regular expressions. For the Text format, we applied regular expression-based pattern matching to identify URL candidates. For the LaTeX format, we extracted URLs from \texttt{.tex} and \texttt{.bbl} files using both regular expressions and the \verb|\url| and \verb|\urladdr| anchors. For HTML files, we extracted URLs enclosed by the \texttt{<a>} tags, and for XML files, we selected elements containing a \texttt{target} attribute (eg., \texttt{<ref target="https://github.com/kermitt2/grobid"/>}). 

After the extraction process, we excluded self-referencing URLs (these are the links that point back to the same paper or its hosting page on arXiv). To evaluate the performance of each format, we extended the ground truth by creating a superset of URLs. This superset includes all URL-like candidates extracted from the Text, LaTeX, HTML, and XML formats in addition to the human-verified URLs from the PDF versions. Extending the set was necessary to accurately compute precision, which requires the total number of extracted URL strings, including both valid and \textbf{invalid} (URL not found in the PDF) URLs.


Framing it as a set-matching problem, we evaluate the URL extraction performance for each individual format and format combination using precision, recall, and F1 score, based on the total number of extracted URLs and the subset identified as valid URLs. We use the following definitions:

\begin{itemize}[labelsep=5pt]
    \item \textbf{Precision} = extracted valid URLs / total extracted URL strings
    \item \textbf{Recall} = extracted valid URLs / total valid URLs (ground truth)
\end{itemize}

Table \ref{tab:performance_combinations} presents the URL extraction performance across individual file formats and their combinations, evaluated against a ground truth of 87 valid URLs.  manually identified from the PDF versions of 10 selected papers.

\renewcommand{\arraystretch}{1.3}
\begin{table}[h]
\centering
\caption{URL extraction performance across format combinations.
P = Precision; R = Recall; V. URLs = Number of valid URLs identified through manual PDF inspection.}
\label{tab:performance_combinations}
\begin{tabular}{p{0.28\columnwidth}p{0.13\columnwidth}p{0.08\columnwidth}p{0.08\columnwidth}p{0.08\columnwidth}}
\hline
\textbf{Format} & \centering \textbf{V. URLs} & \centering \textbf{P} & \centering \textbf{R} & {\centering \textbf{F1}} \\ \hline
Text & \centering 22 & 0.42 & 0.25 & 0.31 \\
LaTeX & \centering 24 & 0.57 & 0.28 & 0.38 \\
HTML & \centering 56 & 0.67 & 0.64 & 0.65 \\
XML & \centering 39 & 1.00 & 0.45 & 0.62 \\ \hline
Text + LaTeX & \centering 34 & 0.41 & 0.39 & 0.40 \\
Text + HTML & \centering 65 & 0.53 & 0.75 & 0.62 \\
Text + XML & \centering 51 & 0.63 & 0.59 & 0.61 \\
LaTeX + HTML & \centering 69 & 0.61 & 0.79 & 0.69 \\
LaTeX + XML & \centering 56 & 0.76 & 0.64 & 0.69 \\
HTML + XML & \centering 60 & 0.69 & 0.69 & 0.69 \\ \hline
Text + LaTeX + HTML & \centering 72 & 0.49 & 0.83 & 0.62 \\
Text + LaTeX + XML & \centering 59 & 0.55 & 0.68 & 0.61 \\
Text + HTML + XML & \centering 69 & 0.55 & 0.79 & 0.65 \\
\textbf{LaTeX + HTML + XML} & \centering \textbf{73} & \textbf{0.62} & \textbf{0.84} & \textbf{0.71} \\ \hline
Text + LaTeX + HTML + XML & \centering 76 & 0.50 & 0.87 & 0.64 \\ \hline
\end{tabular}
\end{table}

First, the results indicate that accurate and complete URL extraction from scholarly papers is a challenging task, and no single format or combination of formats achieves a perfect or even nearly perfect result. Among single formats, HTML achieved the highest F1 score (0.65), with a balanced precision (0.67) and recall (0.64), highlighting its comparative effectiveness for URL extraction. XML exhibited perfect precision (1.00) but only moderate recall (0.45), indicating that while all extracted URLs were valid, it missed many that were present in the PDF files. LaTeX and Text formats showed relatively lower recall (0.28 and 0.25 respectively), with LaTeX outperforming Text slightly in terms of precision (0.57 vs. 0.42).

Combining multiple formats substantially improved recall and the overall F1 scores. For example, the LaTeX + HTML combination achieved an F1 score of 0.69, with a high recall (0.79) and a precision (0.61). Notably, combining three formats (LaTeX + HTML + XML) yielded the highest F1 score (0.71) and recall (0.84). The four-format combination (Text + LaTeX + HTML + XML) achieved the highest recall (0.87) but a slightly reduced F1 score (0.64), primarily due to a lower precision (0.50) caused by an increase in non-valid URL-like strings from the Text format.

The results underscore the importance of leveraging multiple formats to improve URL extraction. While single formats can offer partial coverage, combining other formats, especially HTML and XML, delivers more comprehensive and accurate extraction.

\begin{figure}[h]
\vspace{-0.1cm}
  \centering
    \setlength{\fboxsep}{2.5pt} 
    \setlength{\fboxrule}{0.0pt} 
  \fbox{\includegraphics[trim=15 15 5 1, clip, width=0.65\linewidth]{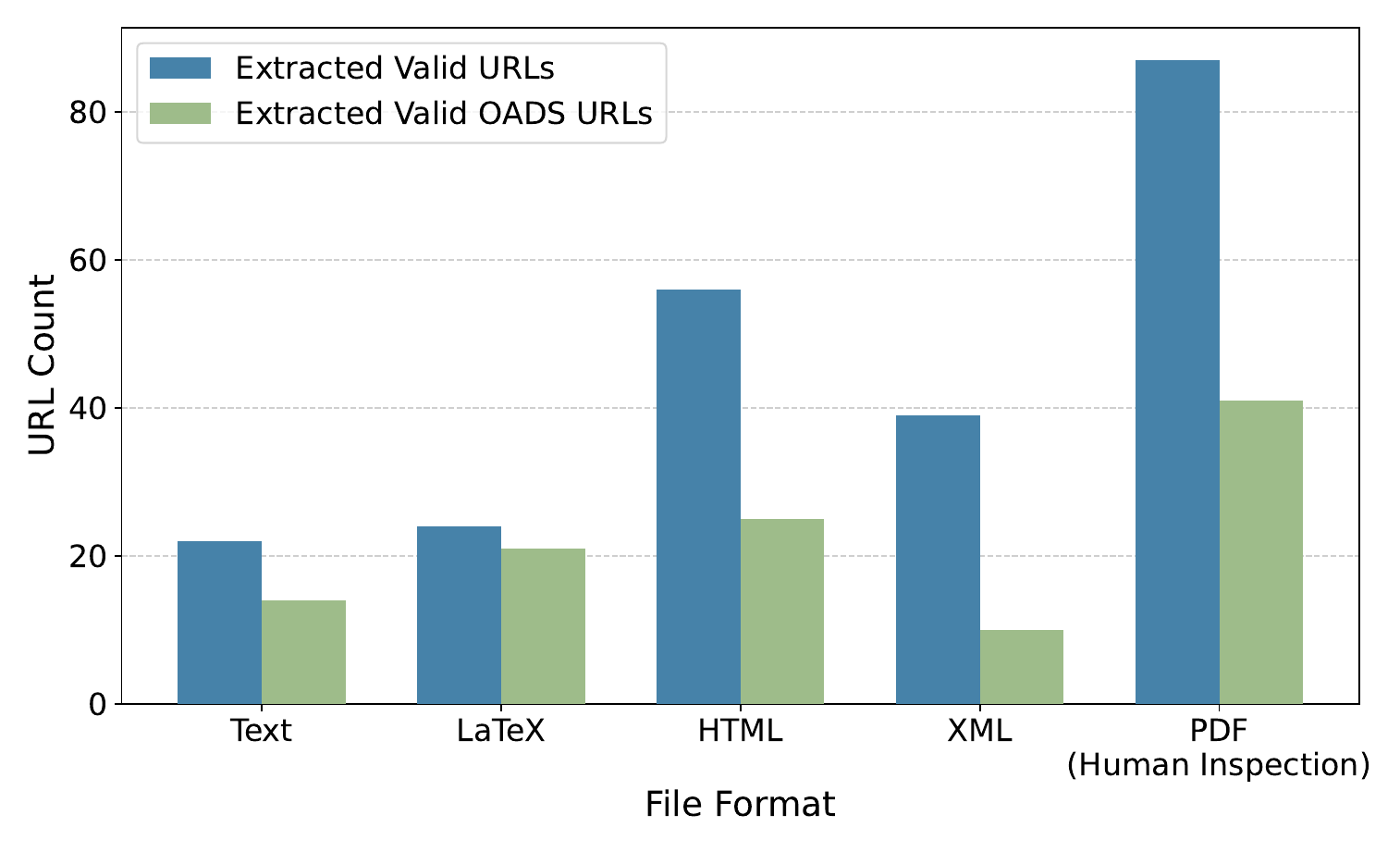}}
    \caption{Number of valid and OADS URLs extracted per file format, compared to human inspection of 10 PDF papers.}
  \label{fig:url_bar}
\vspace{-0.3cm}
\end{figure}

\begin{figure}[h]
\vspace{-0.1cm}
  \centering
    \setlength{\fboxsep}{2.5pt} 
    \setlength{\fboxrule}{0.2pt} 
    \fbox{\includegraphics[trim=45 20 45 20, clip, width=0.65\linewidth]{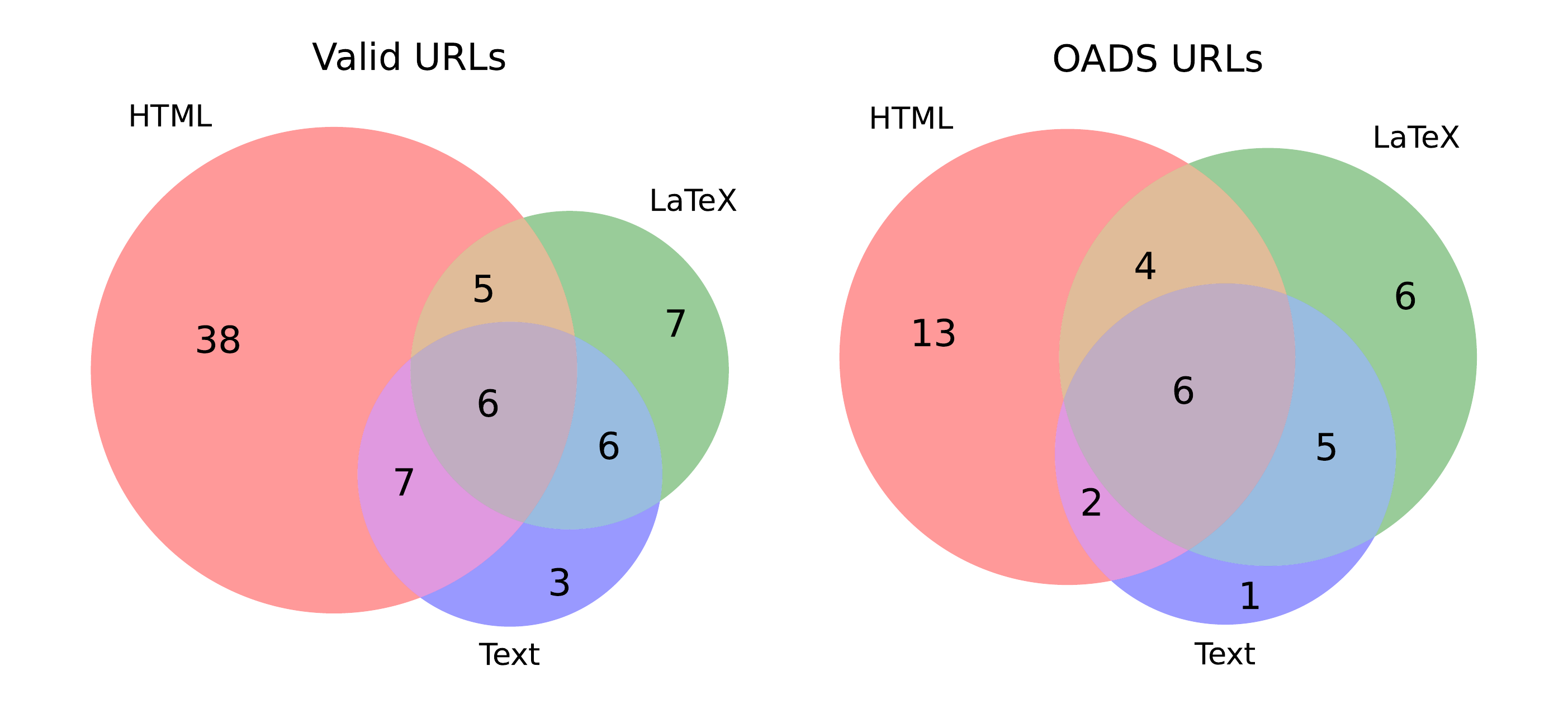}}
    \caption{Overlap of valid and OADS URLs across Text, LaTeX, and HTML formats for the 10 manually annotated papers.}
  \label{fig:url_venn}
\vspace{-0.3cm}
\end{figure}

\begin{figure*}[h]
\vspace{-0.1cm}
  \centering
    \setlength{\fboxsep}{0.5pt} 
    \setlength{\fboxrule}{0.0pt} 
  \fbox{\includegraphics[trim=1 1 1 1, clip, width=1.0\linewidth]{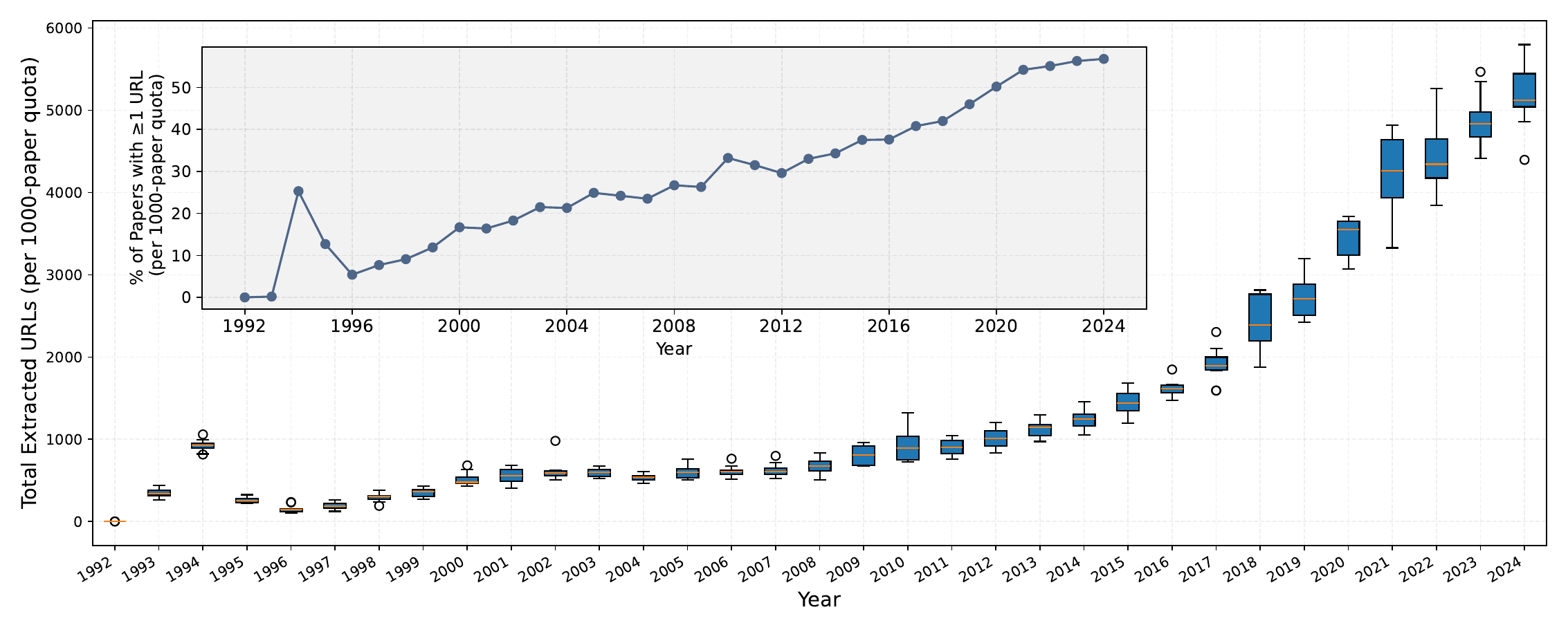}}
  \caption{ A composite distribution of extracted URLs from arXiv papers. The boxplot for each year is obtained by 10 random draws of 1,000 papers with replacement from all papers published in that year. Inset: Percentage of papers containing at least one URL, based on a single random sample of 1,000 papers per year.}
  \label{fig:url_temporal_trends}
\vspace{-0.6cm}
\end{figure*}


\subsection{OADS URLs}
The URLs extracted encompass a wide variety of URLs. Analyzing them, we focus on a particular type of URLs that link to open-access datasets and software (OADS) \cite{lamia-oads}. These URLs have shown to be important for reproducing the findings reported in research papers in AI \cite{kenny-icdar-2023}. To this end, we manually labeled the 87 valid URLs extracted from the PDFs and found that 41 of them are OADS URLs. Figure \ref{fig:url_bar} displays the number of valid OADS URLs and all valid URLs extracted for each format. Among all formats, HTML yielded the highest number of OADS URLs (25) but LaTeX yielded the highest fraction of OADS URLs (87.5\%). The Text format yields a low OADS URL extraction number (14) but a decent extraction rate (63.6\%). We compare the overlap of valid and OADS URL availability across these three formats in Figure \ref{fig:url_venn}.

As shown in Figure \ref{fig:url_venn}, several URLs were extracted from all three formats, but each format also included several URLs not extracted in the others. Figure \ref{fig:url_venn} further shows that relying on a single format can miss a significant number of URLs when they are extracted using a regular expression-based method. Using multiple formats increases the chance toward a more complete set of URLs. A similar conclusion holds for OADS URLs. In particular, extraction from HTML format yields the highest number of valid URLs and OADS URLs.  


\section{Temporal Trends in arXiv URLs}

Understanding how scholars reference web-based resources over time is crucial for designing reliable preservation strategies and improving long-term access to research artifacts. As open science increasingly relies on external datasets, software, and repositories linked via URLs, examining historical patterns of URL usage helps identify shifting practices and emerging dependencies. Here we present a longitudinal analysis of URLs in 2.3 million arXiv papers to reveal how linking behaviors have evolved over three decades. As a pilot study, we built a subset of arXiv papers by randomly sampling 1000 arXiv papers per year from 1992 to 2024, resulting in a total of $\sim$33,000 papers.

To account for sampling variability and ensure robust yearly estimates, we repeated the sampling process ten times. Each time, we obtain the number of URLs extracted for each year. By combining the data collected in 10 times, we produce a composite URL count distribution in Figure \ref{fig:url_temporal_trends}. We extracted URLs from each paper's Text format using regular expression-based pattern matching. Although the Text format did not yield the most accurate or complete URL extractions compared to other formats, we included it in our study because it was readily and consistently available across the full arXiv corpus at the time of analysis. It allowed us to assess the trend with minimal preprocessing.

Figure \ref{fig:url_temporal_trends} inset presents the percentage of papers containing at least one URL over time from the 1000 selected paper samples per year. The adoption of URLs in arXiv papers has increased significantly over the past three decades. In 1993, fewer than 0.02\% of sampled papers included even a single URL, reflecting the limited use of the internet in scholarly communication at the time. Adoption remained below 10\% through the mid-1990s (except for 1994), but began rising steadily after 1998, coinciding with the broader availability of web browsers and growing reliance on online resources in academic research \cite{History-of-the-Web, De_Groote2014-xx}.


\section{Discussion}


Our pilot study indicates that no single file format provides accurate and complete coverage of URLs found in the original scholarly papers. Indeed, the format of the input document strongly affects the URL extraction. Structured formats such as HTML and LaTeX usually yield better results because they keep the content clean and machine-readable. HTML includes semantic tags that separate URLs from other text, and LaTeX preserves the original content without layout issues. In contrast, text extracted from PDFs often deform URLs through formatting complexities such as line breaks, hyphenation, or embedded graphics. We found that no single format captures OADS links consistently. Combining multiple formats, particularly HTML, XML, and LaTeX, improves coverage by $\sim$39\% to $\sim$51\% compared with relying on only the Text format. However, combining formats can also increase false positives. One solution is to filter or score links based on confidence. 


Our temporal analysis reveals an accelerating dependency on external digital resources in scientific research. The increase in papers containing linked resources, from only 0.02\% in the early 1990s to approximately 55\% in recent years, marks a fundamental shift in scholarly communication and knowledge sharing. The notable spike in 1994, where around 25\% of papers contained URLs, stood out from the much lower rates in surrounding years. At first, this seemed like an artifact in the data, but further investigation confirmed it was real and reflected the early growth of the World Wide Web.
By late 1993, over 500 web servers were active \cite{cern_web_history}, and several papers from 1994, particularly in Computer Science and Physics, cited these early web resources, contributing to a temporary but genuine increase in URL usage. This finding illustrates how individual researchers or research groups can drive the adoption of new technologies before broader community acceptance occurs. The 1994 anomaly represents early adopters experimenting with web-based resource sharing, indicating the widespread adoption that would follow several years later. 

Our study faces several limitations that potentially affect the generalizability of our findings. The first is the small sample size because not all formats are available for most papers.  
Additionally, in the temporal trend analysis, the results are based on URLs extracted from Text files. Further experiments using other input formats should be conducted to validate and confirm these findings.

\section{Conclusion and Future work}

In this study, we examined URL extraction performance across multiple arXiv file formats using a pilot dataset, revealing clear differences in extraction accuracy and completeness. Markup formats such as HTML and XML generally provide more accurate and comprehensive URL extraction compared to Text format. Our results highlight that relying on a single file format risks missing important links, especially those related to open-access datasets and software. Combining formats improves coverage but requires attention to the unique challenges each format presents. Our temporal analysis reveals how referencing web-based resources in scholarly papers has steadily increased over the past thirty years. 

Several research directions emerge from our findings that could substantially advance scholarly preservation efforts. Scaling human annotation to more papers would provide more robust ground truth data and enable detailed error analysis across different research domains. Expanding our analysis to include S2ORC and PubMed would provide insights across scholarly repositories in a broader range of domains. 



\bibliographystyle{unsrt}  
\bibliography{references}  

\clearpage

\end{document}